# Grain Boundary Softening from Stress Assisted Helium Cavity Coalescence in Ultrafine-Grained Tungsten

W. Streit Cunningham[1], Yang Zhang[1], Spencer L. Thomas[1], Osman El-Atwani[2], Yongqiang Wang[2], and Jason R. Trelewicz[1,3,*]

[1]Department of Materials Science and Chemical Engineering, Stony Brook University, Stony Brook, NY 11794
[2]Materials Science and Technology Division, Los Alamos National Laboratory, Los Alamos, NM 87545
[3]Institute for Advanced Computational Science, Stony Brook University, Stony Brook, NY 11794

*Corresponding Author at Stony Brook University: jason.trelewicz@stonybrook.edu

## Abstract

The formation of helium cavities in coarse-grained materials produces hardening proportional to the number density and size of the cavities and due to the interaction of dislocations with intragranular helium defects. In nanostructured metals containing a high density of interfacial sinks, preferential cavity formation on the grain boundaries instead produces softening and often attributed to enhanced interfacial plasticity. Employing two grades of ultrafine-grained tungsten, we explore this effect using targeted implantation studies to map cavity evolution as a function of the irradiation conditions and quantify its impact on the mechanical response through nanoindentation. Softening is reported at implantation temperatures above the threshold for preferential grain boundary cavity formation but at a sufficiently low fluence prior to the growth of intragranular cavities. Collective changes in the mean cavity size, density, and morphology beneath a residual impression on an implanted surface indicate that cavity coalescence accompanied the reduction in hardness. Complementary atomistic simulations demonstrate that, in tungsten grain structures exhibiting softening, grain boundary bubble coalescence is driven by stress concentrations that further act to localize strain in the grain boundaries through cooperative deformation processes involving local atomic shuffling and sliding, dislocation emission, and even the nucleation of unstable twinning events.

1. Introduction

A significant body of literature has focused on nanocrystalline metals as radiation-tolerant materials due to the high density of interfaces potentially leading to enhanced annihilation of irradiated-induced defects through the absorption of interstitials at interfaces [1, 2]. At low homologous temperatures, biased diffusion of interstitials to grain boundaries promotes elevated vacancy concentrations within the grain matrix but with recombination in the immediate vicinity of the interfaces and subsequent formation of a denuded zone [3, 4]. Four stages of defect annihilation in nanocrystalline metals consistent with this mechanism have been proposed: stages I and II where defect density increases with dose, stage III where the rate of defect production and annihilation are similar, and stage IV where defect annihilation becomes the dominant behavior [5]. One implication of stage IV is that nanocrystalline metals should exhibit recovery following a saturation dose, which has been noted in several systems based on an observed reduction in the total defect density with increasing fluence [5-10]. An intrinsic mechanism for recovery following a critical saturation dose is unique to nanostructured materials where the microstructural and diffusion length scales are comparable [11-13], thereby providing a pathway for tuning a material's radiation tolerance via defect sink engineering to suppress classical degradation phenomena such as irradiation embrittlement [14, 15] and void swelling [16, 17].

Control over defect sinks in engineered microstructures has been a large focus of the materials development efforts in structural nuclear materials [13, 18, 19]. Specific to first-wall materials in a fusion reactor are extreme environments containing fast 14-MeV neutrons, high deuterium, tritium, and helium (He) particle fluxes ($10^{23}$-$10^{24}$ m$^{-2}$s$^{-1}$), and excessive thermal loads (5-20 MW/m$^2$) [13]. Displacement damage from the 14 MeV neutron spectrum will be accompanied by the formation of transmutation products such as He gas (10-12 appm He/dpa in steels) [20, 21], which when combined with the He flux from the plasma, are expected to generate He concentrations of over 2,000 appm over the life-time of the material [13]. The complete insolubility of He in most metals typically results in the formation of He clusters, He-induced vacancies, and He-vacancy complexes that subsequently coalesce into cavities; the amount of He within the cavity determines whether the cavity is a bubble (containing He) or a void (free of He) [22, 23]. The nucleation of cavities is exacerbated at higher temperatures due to the high binding energy of He with vacancies, which reduces the critical radius for bias-driven void growth [24]. The combination of this behavior with the affinity of He for microstructural sinks has been shown to result in significant heterogeneous cavity nucleation at grain boundaries and dislocations [16, 22, 25, 26], promoting higher swelling rates and early mechanical failure through embrittlement [27-29].

The acceleration of void swelling by He gas, combined with dislocation loop damage, makes W susceptible to embrittlement even at low doses (<0.3 dpa) [30, 31] and also degrades its properties as evidenced by reductions in thermal conductivity and corresponding increases in the already high ductile-to-brittle transition temperature (DBTT) [32, 33]. An additional consideration for W as a plasma facing material is the formation of nanofuzz under a high flux of low energy ions, which in some literature has been attributed to the bursting of high-pressure subsurface bubbles [34-36]. Degradation of the surface from nanofuzz combined with the reduction in thermal conductivity under neutron irradiation also raises concerns about W exceeding its recrystallization temperature [37, 38]. To address these intrinsic limitations of W for the fusion environment, tailored nanostructures have gained interest over the past decade, and recent studies suggest that improvements in certain performance metrics relevant to fusion can be realized through the addition of a high density of interfaces [10, 39, 40]. Combined with enhanced thermal stability exhibited by a growing number of nanostructured alloys across different alloy systems [41-45], the high interfacial density has also been shown to affect radiation tolerance, largely manifested through increased dislocation annihilation at grain boundaries relative to coarse-grained W [5, 10, 46]. However, its implications for He cavity formation and the associated changes in mechanical properties are still an active area of research, especially given observations of softening under certain He irradiation conditions where hardening has been traditionally reported [22, 47, 48].

Prior work has demonstrated that He cavity microstructures are heavily dependent on temperature and fluence with preferential accumulation along grain boundaries as discussed above, but also at other sinks such as dislocations, and/or precipitates at most temperatures and doses [49, 50]. Grain boundary cavity formation is profuse in nanocrystalline materials due to their intrinsically high interfacial sink density [11], whereas in coarse grained materials, cavities are more uniformly distributed with the formation of a bubble superlattice at low temperatures/doses (T≈0.35 $T_m$) [51, 52]. Continuous bubble growth beyond a critical size (typically ~ 10 nm) results in a "bubble-to-void" transition due to the diffusion and fast absorption of vacancies [2, 53] that is largely deleterious due to its association with increased embrittlement [54-57]. Furthermore, it has been shown in polycrystalline Cu [58, 59], Fe [60], Ni [61], and W [47, 62] that the presence of He cavities in the grain matrix inhibits dislocation motion and results in an increase in hardness with increasing He concentration as described by the Orowan model [63, 64]:

$$\Delta\tau = \alpha G b \sqrt{Nd} \quad (1)$$

In Eq. 1, the change in shear stress, Δτ, is a function of the defect barrier strength $\alpha$, the shear modulus $G$, the magnitude of the burgers vector b, and the number density and size of defects, $N$ and $d$, respectively. However, as noted above, a recent study by the authors demonstrated the existence of softening at low He concentrations in W, which was discussed in the context of preferential He cavity formation at grain boundaries [22]. Given the potential for interfaces to serve as effective defect sinks, it is critical to develop a fundamental understanding of mechanisms governing He cavity evolution at grain boundaries along with its consequences for mechanical behavior.

In this study, the influence of the helium cavity distributions on the mechanical behavior of fine-grained tungsten is explored using targeted implantation studies. Nanoindentation hardness and complex modulus are mapped as a function of irradiation temperature and fluence in two different tungsten microstructures. Scaling behavior is explained based on the helium cavity microstructure with a focus on the distribution of cavities between the grain matrix and boundaries as quantified through transmission electron microscopy (TEM). Based on measurements of cavity evolution indicating that coalescence accompanies softening, a complementary atomistic simulation study was performed to map the deformation behavior of W grain structures containing grain boundary He bubbles. We establish correlations between bubble evolution and the flow behavior that are consistent with experiments and demonstrate stress-assisted bubble coalescence promotes softening as a result of a shift to grain boundary mediated processes such as local atomic shuffling and sliding, dislocation emission, and even the potential for the nucleation of unstable localization events such as deformation twinning.

## 2. Materials and methods

### 2.1. Experimental

Two different ultrafine-grained W samples were employed in this study: a commercial-grade ultrafine-grained W produced via cold rolling provided by Alfa Aesar (denoted AA-W) and severe plastically deformed W containing a mixture of ultrafine and nanocrystalline grains (denoted SPD-W). Additional details about the AA-W and SPD-W microstructures are provided in Refs. [65] and [66], respectively. All samples were annealed for one hour at 950°C (representing the highest temperature irradiation condition) to stabilize the microstructure for irradiation at elevated temperatures with the surface subsequently polished to a mirror finish prior to irradiation.

. Samples were implanted under several conditions to map both temperature and fluence effects in the different grade microstructures: (i) 800°C, AA-W was subjected to 500 keV He$^+$ ions in a 3 MV NEC tandem accelerator, located at the Ion Beam Materials Lab (IBML) at Los Alamos National Laboratory (LANL), to a fluence range of 1 x 10$^{15}$ - 4 x 10$^{16}$ ions/cm$^2$, and (ii) 950°C, both AA-W and SPD-W were subjected to 150 keV He$^+$ ions in a 200 kV Varian ion implanter, also located at the IBML, to a fluence range of 1 x 10$^{15}$ - 1 x 10$^{17}$ ions/cm$^2$. Temperatures were selected according to the temperature threshold for biased grain boundary bubble formation described in Ref. [49] due to diffusion of He-V complexes at 950°C relative to 800°C where only interstitials and vacancies are mobile [67]. Representative damage and He concentration profiles were generated using a 70 eV displacement energy [68] in the Stopping Range of Ions in Matter (SRIM) code [69] and are shown in Figure 1a with the estimated peak He concentrations from SRIM provided as a function of fluence in Figure 1b.

Following implantation, TEM samples were prepared from the bulk material through a typical focused ion beam (FIB) lift-out procedure using two separate FEI Helios Nanolab 600 DualBeam FIBs, one located at the Center for Functional Nanomaterials (CFN) at Brookhaven National Laboratory and the other at the Electron Microscopy Lab (EML) at LANL. Characterization of the cavity microstructure was performed using two different TEMs: a 300 keV FEI Tecnai F30 located in the EML at LANL and a 200 keV JEOL JEM 2100F located at the CFN. Quantification of the He cavity microstructures was accomplished using ImageJ and Adobe

Photoshop software. Cavity sizes and densities were measured across numerous regions for each sample and averaged to obtain representative values for each irradiation condition. To obtain cavity measurements at grain boundaries, a modified analytical procedure was employed where measurements were only acquired from cavities at heavily inclined boundaries. As TEM micrographs are akin to a 2D projection of a 3D space, the influence of the grain boundary tilt on measured cavity sizes can be compensated for given the total boundary length and an assumed sample thickness of 100 nm. Furthermore, as there is no precise methodology for determining grain boundary width from basic bright-field TEM micrographs, areal grain boundary cavity densities are only provided herein (cavities/nm$^2$).

Nanomechanical properties including the indentation hardness and complex modulus were mapped as a function of contact depth using a Bruker TI980 Triboindenter coupled to a dynamic mechanical analysis (DMA) transducer for continuous stiffness measurement (CSM). Experiments were conducted at room temperature with a 150 nm Berkovich probe, which was placed in contact with the specimen surface under a load of 2 μN for a minimum of one hour prior to testing to minimize the influence of thermal drift. Indentation strain rate was held constant at 0.5 s$^{-1}$ with experiments performed to a maximum load of 10 mN under the application of a dynamic oscillation force such that the displacement amplitude was maintained between 1-2 nm and frequency fixed at 200 Hz. Instrumental drift was characterized during the linear unloading segment using a 10 s hold at 10% of the maximum load. A minimum of 30 indents with negligible drift were acquired per sample to produce statistically significant measurements analyzed via the Oliver and Pharr method [70] using a tip area function calibrated on fused silica.

*2.2. Simulations*

Three nanocrystalline W grain structures with varying He concentrations were employed to explore mechanical property scaling and its mechanistic underpinnings in the presence of grain boundaries containing He bubbles. To maintain consistency through the simulations, all three structures shared an identical grain structure containing 25 grains with an average grain size of 15 nm inside a cubic simulation box with a 35.5 nm side length and a total volume of ~44,700 nm$^3$ (corresponding to ~2.8 million atoms). The grain size distribution was optimized through a Monte Carlo (MC) procedure to eliminate any extremely small (<5 nm) or large (>20 nm) grains as well as grains with irregular shapes (e.g., polygons with a small number of sides). The introduction of He was accomplished through random site occupation on the BCC W lattice in concentrations of 1 at.% He and 5 at.% He, corresponding roughly to the He concentrations at fluences of 1 x 10$^{16}$ ions/cm$^2$ and 1 x 10$^{17}$ ions/cm$^2$, respectively (as estimated from the peak He concentrations in Figure 1b). This method allowed for the total number of atoms to remain constant while not assuming a predefined He to vacancy ratio to maintain a constant number of atoms in the system. The hybrid MC-molecular dynamics (MD) scheme described in Ref. [71] was then employed for achieving an energy-minimized state where He was redistributed through 100 MC steps seeded in 100,000 MD relaxation steps using the Large-scale Atomic/Molecular Massively Parallel Simulator (LAMMPS) platform [72]. The energy-minimized and relaxed simulation structures are referred to henceforth as Pure W, W-1%He, and W-5%He. Analysis and visualization images were prepared with the assistance of the OVITO software [73].

The interatomic potential selected for this study was specifically designed for He bubbles in W [74]. Energy minimization employed a final relative energy convergence of 10$^{-12}$ with periodic boundary conditions applied in all directions. The isothermal-isobaric ensemble was employed during relaxation via simulated annealing that involved heating to 377°C at a rate of 0.1 K/ps where the structure was held for 1 ns prior to cooling to 27°C at the same rate. Finally, an additional relaxation step was performed for 1 ns at 27°C again using the isothermal-isobaric ensemble to achieve a zero-pressure condition along all directions. Uniaxial tensile simulations were performed on the LAMMPS platform at a fixed strain rate of $10^7$ s$^{-1}$. The strain rate was selected to sample dislocation plasticity based on results from trial simulations at three different strain rates of $10^7$ s$^{-1}$, $10^8$ s$^{-1}$, and $10^9$ s$^{-1}$.

## 3. Cavity formation and distribution in the tungsten microstructure

The pristine microstructures of the AA-W sample exhibited elongated ultrafine grains with a {001}<110> fiber texture while the SPD-W microstructures contained a bimodal distribution of nanocrystalline and ultrafine grains with a random texture; further details on their characteristics including quantitative grain size analysis, texture, and grain boundary character distribution, etc. can be found in Refs. [65, 66]. The different cavity microstructures that developed as a function of irradiation temperature and fluence are depicted in Figure 2 with bright-field TEM micrographs for AA-W in panels (a,b) and SPD-W in panel (c). Qualitative examination of the AA-W microstructure at 800°C and at the lowest fluence (1 x $10^{15}$ ions/cm$^2$) in the upper panel of Figure 2a reveals the presence of small approximately circular cavities distributed randomly throughout the microstructure. An increase in fluence to 4 x $10^{16}$ ions/cm$^2$ was accompanied by a marked increase in the cavity density but with no apparent increase in cavity size, as evidenced in the lower panel of Figure 2a. Different behavior was observed when irradiating at 950°C where the small approximately circular cavities homogeneously distributed in the microstructure at the lower fluence of 1 x $10^{15}$ ions/cm$^2$ in the upper micrograph of Figure 2b transitioned to a bimodal distribution of cavities at the higher fluence of 1 x $10^{17}$ ions/cm$^2$ as shown in the lower micrograph of Figure 2b. Larger cavities were biased to the grain boundaries with comparatively smaller faceted cavities forming in the grain matrix. Qualitatively, the microstructures of SPD-W resulting from He implantation at 950°C shown in Figure 2c exhibited cavity distributions akin to AA-W. At the lowest fluence of 1 x $10^{15}$ ions/cm$^2$, small cavities were randomly distributed throughout the microstructure in the upper micrograph of Figure 2c, whereas at the highest fluence of 1 x $10^{17}$ ions/cm$^2$, larger faceted cavities formed throughout the grain matrix and boundaries with the largest cavities located at the grain boundaries. The transition to faceted cavity morphologies at higher irradiation temperatures has been widely observed in He implanted materials [75, 76] and typically attributed to the crystallographic constraints imposed by the lattice symmetries [77]. Furthermore, the transition to a bimodal cavity distribution in both W grades is consistent with prior observations of a temperature threshold for preferential grain boundary cavity formation [49] as is the presence of back dot damage accompanying the cavity microstructure and indicative of defect clustering and loop formation at both fluences [78].

Cavity size distributions quantified from the TEM micrographs are shown in Figure 3 for AA-W and SPD-W under all implantation conditions and delineated for the grain matrix and grain boundary regions. In AA-W implanted at 800°C, the average cavity size at a fluence of 1 x $10^{15}$ ions/cm$^2$ was nominally equivalent in the matrix and grain boundaries with only minimal evolution in the grain boundaries from approximately 2 to 4 nm with increasing fluence to 4 x $10^{16}$ ions/cm$^2$. The increase in temperature to 950°C led to a small increase in the average cavity size in the grain matrix at a fluence of 1 x $10^{15}$ ions/cm$^2$; however, increasing the fluence to 1 x $10^{17}$ ions/cm$^2$ at this temperature led to coarsening of the grain matrix cavities by approximately 5X in both AA-W and SPD-W, demonstrating consistent intragranular bubble behavior in both grades of W. In the grain boundaries at the lowest fluence of 1 x $10^{15}$ ions/cm$^2$, cavities were marginally larger than those in the matrix for the AA-W implanted at 950°C while the SPD-W exhibited similar size distributions between the matrix and grain boundary under these implantation conditions. At the largest fluence of 1 x $10^{17}$ ions/cm$^2$, significantly larger cavities formed in the grain boundaries in both grades of W as evidenced by the distributions shifting to larger diameters and consistent with the temperature threshold for biased grain boundary cavity formation in W [49].

A quantitative analysis of the cavity distributions is provided in Figure 4a for the lowest and highest fluence conditions at each implantation temperature. The grain matrix and boundaries exhibited comparable cavity densities at both fluences in the AA-W sample implanted at 800°C, which increased sharply with increasing fluence. With the size distributions being similar under these conditions from Figure 3, the increase in fluence at 800°C promoted cavity formation rather than growth and coalescence. This increase in cavity density was less pronounced at 950°C over two decades of fluence for both grades of W and with a marked disparity between the grain matrix and boundaries. The densities were consistently less in the grain boundaries, which was exacerbated for the largest fluence condition but with a distinguishing feature between the AA-W and SPD-W. At both fluences, the disparity between the matrix and grain boundary cavity densities were more pronounced in AA-W relative to SPD-W. The grain boundary cavities were also larger in the AA-W sample across the two decades of fluence at 950°C. Collectively, these results indicate that the grain boundaries more strongly influence the accumulation of He in AA-W, which is consistent with its high fraction of nanocrystalline grains parallel to the surface in the textured microstructure [65] providing favorable diffusion pathways for He to accumulate in the grain boundaries.

The total damage as a function of implantation temperature and fluence can be represented by the cavity induced swelling as estimated from the cavity densities and size. For the grain matrix, the average volumetric swelling ($\Delta V/V$) was calculated for each implantation condition:

$$\frac{\Delta V}{V} = \frac{\left(\frac{4\pi r^3}{3}\right)\rho}{1-\left(\frac{4\pi r^3}{3}\right)\rho} \qquad (2)$$

where $r$ is the average cavity radius and $\rho$ is the average cavity density [58]. Because the grain boundary width cannot be precisely determined from bright-field TEM micrographs, grain boundary swelling was quantified using the average areal swelling ($\Delta A/A$):

$$\frac{\Delta A}{A} = \frac{(\pi r^2)\sigma}{1-(\pi r^2)\sigma} \qquad (3)$$

where $\sigma$ is the average cavity density in the grain boundary. On this basis, we note that direct quantitative comparisons of the swelling magnitude between the matrix and boundaries are not appropriate given the difference in areal versus volumetric estimations. Average values for matrix volumetric and grain boundary areal swelling are provided in Table 1 and Table 2, respectively, and plotted in Figure 4b. At the lowest fluence across both temperatures in AA-W and SPD-W, average volumetric swelling within the grain matrix remained below 0.1% and increased monotonically with fluence for both temperatures. This scaling of the grain matrix swelling with fluence was exacerbated by temperature (greater than one order of magnitude), as evidenced in the 950°C implantation data. Overall trends in the grain boundary swelling follow the grain matrix where grain boundary swelling increased significantly with fluence in both microstructures with some variation in temperature, particularly at the highest fluence. Specifically, the AA-W samples exhibited the most severe swelling at the highest fluence of 1 x $10^{17}$ ions/cm² and consistent with a similar density of larger cavities formed in the grain boundaries relative to the SPD-W.

**Table 1.** Average cavity diameter, density, and volumetric swelling for grain matrix cavities for all conditions.

| Microstructure | Temp. (°C) | Fluence ($10^{15}$ cm$^{-2}$) | Avg. Cavity Diameter (nm) | Avg. Cavity Density ($10^{-3}$ nm$^{-2}$) | Avg. Volumetric Swelling ($10^{-4}$) |
|---|---|---|---|---|---|
| AA | 800 | 1 | 2.4 ± 0.6 | 2.9 ± 1.3 | 2.3 ± 0.5 |
|  |  | 40 | 2.7 ± 0.5 | 17.3 ± 2.5 | 18.1 ± 3.4 |
|  | 950 | 1 | 3.7 ± 1.9 | 2.7 ± 1.3 | 7.3 ± 1.0 |
|  |  | 100 | 10.2 ± 3.2 | 9.5 ± 2.0 | 548 ± 34 |
| SPD | 950 | 1 | 3.0 ± 1.1 | 3.1 ± 1.1 | 4.2 ± 0.7 |
|  |  | 100 | 9.6 ± 3.2 | 6.6 ± 2.0 | 313 ± 34 |

**Table 2.** Average cavity diameter, density, and areal swelling for grain boundary cavities for all conditions.

| Microstructure | Temp. (°C) | Fluence ($10^{15}$ cm$^{-2}$) | Avg. Cavity Diameter (nm) | Avg. Cavity Density ($10^{-3}$ nm$^{-2}$) | Avg. Areal Swelling ($10^{-4}$) |
|---|---|---|---|---|---|
| AA | 800 | 1 | 1.7 ± 0.2 | 3.5 ± 1.7 | 4.9 ± 0.5 |
|  |  | 40 | 4.1 ± 1.9 | 15.9 ± 3.3 | 10.0 ± 0.8 |
|  | 950 | 1 | 5.6 ± 1.9 | 0.8 ± 0.2 | 7.1 ± 0.6 |
|  |  | 100 | 26.8 ± 9.1 | 0.9 ± 0.2 | 1040 ± 65 |
| SPD | 950 | 1 | 2.9 ± 0.8 | 1.9 ± 0.3 | 2.3 ± .008 |
|  |  | 100 | 17.7 ± 7.8 | 1.1 ± 0.4 | 317 ± 92 |

The reported sizes and density distributions of He cavities agree with prior measurements in several ultrafine-grained and nanocrystalline systems. At 800°C, a high density of small (< 5 nm) cavities was measured in AA-W; similar distributions at temperatures below 850°C have been reported by the authors on ultrafine-grained W [22] and in various additional works in W [49, 79-81] as well as other metals such as Cu [58, 82] and Fe/steels [83, 84]. This behavior is expected given the implantation temperature, as the low He-vacancy cluster mobility at 800°C impedes the formation of larger He cavities [49]. The inverse behavior is expected at higher temperatures due to enhanced He-vacancy cluster migration and is evident in the size distributions of cavities across both W microstructures in this work (Figure 3) and others [22, 49, 82, 85, 86].

## 4. Nanomechanical properties and their dependence on cavity microstructure

Given the inherent depth dependence of the cavity microstructures formed under the different implantation conditions (i.e., as largely dictated by the ion energies but also included by implantation temperature and fluence), appropriate depth parameters for the nanoindentation experiments were ascertained by mapping the cavity distributions and sizes as function of depth from the implantation surface. Figure 5 show cross-sectional bright-field TEM micrographs of the 1 x $10^{15}$ ions/cm$^2$ and 1 x $10^{17}$ ions/cm$^2$ He implanted SPD-W samples, respectively. Superimposed on each micrograph is a depth scale divided into 100 nm regions used for quantitative mapping of cavity distributions. Below each micrograph are the corresponding plots of average cavity density, size (area), and estimated He content as a function of depth from the implantation surface determined by:

$$P = \frac{2\gamma}{r} \quad (4)$$

where P is the pressure, r is the cavity radius, and $\gamma$ is the cavity surface tension. This estimation assumes the cavity pressure is nominally at equilibrium, following the procedure outlined in Ref. [58]. The depth dependent He concentrations from SRIM are overlaid on the He content profiles calculated from the TEM micrographs. We note that these calculations represent rough estimates, and determination of the He concentrations within each cavity would require analytical microscopy techniques that are beyond the scope of this work. Comparatively, cavities formed deeper into the sample with larger sizes at the peak of the distribution under the highest fluence. However, at both fluences, the estimated He content closely matches the He distribution from SRIM, indicative of minimal He diffusion into the sample. Furthermore, this agreement between both distributions suggests that the ideal nanoindentation depth for sampling He effects on the mechanical behavior is approximately 350 nm, which corresponds to contact depths of 120 nm given that the plastic zone is typically assumed to extended three times beyond the contact depth [87]. Note, however, that the plastic zone depth can vary depending on the local microstructure with experiments demonstrating that it can extend from two to nominally five times the indentation depth for BCC metals [87].

Among the several challenges and limitations associated with nanoindentation of irradiated surfaces [88], the issue with the most severe implications for hardness measurement via nanoindentation is biased sampling of the pristine region below the implanted surface due to the inhomogeneous dose profile produced by He ion implantation [89]. While our experiments were designed to access the implanted regions, we also directly confirmed that the depths probed through nanoindentation appropriately sampled the cavity microstructures. A TEM cross-section of an indent from the SPD-W sample implanted to a fluence of 1 x $10^{17}$ ions/cm$^2$ was prepared through a typical FIB lift-out procedure with the resulting bright-field micrograph of the region surrounding the residual impression shown in Figure 6a. The indent had an approximate maximum contact depth of 140 nm, and average cavity density and area were measured for each subsequent 100 nm region below the impression surface. An obvious plastic zone was not immediately apparent below the indent and consistent with other ion implanted fine-grained microstructures [90, 91] where the significant contrast arising from high defect densities and residual strain led to difficulties in discerning the extent of the plastic zone. Depth dependent volumetric swelling trends for the as-implanted and deformed regions are provided in Figure 6b with a superimposed He concentration profile from SRIM. The average cavity density and areas used to calculate swelling are included in Figure 6c. The volumetric swelling profile for the as-implanted region closely follows the predicted He distribution produced through SRIM, peaking at nominally 350 nm. A significant reduction in the volumetric swelling was observed upon nanoindentation that derived from the reduction in the peak cavity density despite the increase in peak cavity area that transpired further below the indented surface.

The changes in the size distribution beneath the residual impression, particularly the reduction in density that accompanied the increase in cavity area, indicate that cavity coalescence transpired within the plastic zone during deformation. To confirm that cavities contributed to the plastic response, we examined their morphological evolution below the indented and as-implanted surfaces in the SPD-W sample irradiated to a fluence of $1 \times 10^{17}$ ions/cm$^2$. The circularity, which is bounded between 0 for highly non-circular shapes and unity for a perfect circle, was employed for this purpose and is shown in Figure 7 as a function of depth from the respective surface through the implanted region. Beneath the as-implanted surface, cavities were approximately circular, with an average circularity of 0.86. In the implanted region beneath the residual impression, the minimum circularity was reduced to approximately 0.6 at a nominal depth of 500 nm, which corresponded to the depth where the cavity area maximized. At larger depths, the cavity circularity increased and eventually converged to the mean value for the as-implanted surface. These observations indicate that the plastic zone expanded into the entirety of the implanted region with cavity coalescence accompanying their deformation.

Through-thickness CSM measurements were performed on all samples with average values for the hardness and modulus extracted from each experiment corresponding to the peak He implantation concentration depths (110 – 130 nm for 150 keV implantations and 250 – 270 nm for 500 keV implantations). Values for hardness and complex modulus as a function of fluence are provided in Figure 8(a,b), respectively, along with the reported ranges for the pristine annealed samples highlighted in gray. In AA-W (Figure 8 top), no discernible change in hardness or complex modulus was observed through fluence for the implantations performed at 800°C.

However, implantation of He at 950°C produced a significant reduction in the measured hardness of the AA-W sample at the lowest fluence, which recovered to the range of the pristine sample with increasing fluence but with a positive slope. The complex modulus exhibited the opposite trend, specifically falling below the pristine range with increasing fluence. The hardness trends in SPD-W at 950°C (Figure 8 bottom) follow a similar pattern to AA-W but with a less drastic reduction in the hardness at low fluence and a more pronounced linear hardening regime across the full fluence range. Additionally, the complex modulus was generally lower at 950°C and consistent with the reduction in modulus that accompanied hardening in the AA-W sample. This increase in hardness as a function of fluence is an understood behavior in irradiated metals [92-94] deriving from the combination of dislocation loops and intragranular He cavities inhibiting dislocation motion (and hence the absence of a plateau since hardening will continue with increasing fluence based on this mechanism). We also note that the hardening trend aligns with prior mechanical measurements on He implanted SPD-W irradiated under different conditions [22, 95] with the increased intragranular cavity size identified as the underlying factor for the measured increased in hardness.

Changes in the estimated He content from TEM as compared with the SRIM concentration profiles suggest a change in cavity pressure that could contribute to the increase in hardness. Total He contents estimated from the SPD-W TEM micrographs in Figure 5 indicate a total He content of $1.6 \times 10^9 \pm 2.3 \times 10^8$ He atoms at a fluence of $1 \times 10^{15}$ ions/cm$^2$ and $1.4 \times 10^{12} \pm 1.7 \times 10^{11}$ He atoms when the fluence increased to $1 \times 10^{17}$ ions/cm$^2$. Count estimates from SRIM are $1.5 \times 10^9$ and $1.5 \times 10^{11}$ He atoms at fluences of $1 \times 10^{15}$ and $1 \times 10^{17}$ ions/cm$^2$, respectively. At the lowest fluence of $1 \times 10^{15}$ ions/cm$^2$, the estimated He content is nominally equivalent with the SRIM results, which suggests that the cavities are bubbles with nominally equilibrium pressure.

Conversely, at the highest fluence of 1 x $10^{17}$ ions/cm$^2$, He count estimates are nine times greater than the SRIM estimates, suggesting that the cavities are larger than expected at this fluence and are therefore underpressurized. However, due to the inhomogeneous cavity size profile at the highest fluence, not all cavities are necessarily underpressurized with those at the tail likely differing from the cavities at the center of the profile. This implied reduction in cavity pressure with increasing fluence matches closely the established bubble-to-void transition [2, 53], where an increase in the bubble radius beyond a critical value often leads to boundless cavity growth through enhanced vacancy capture. MD simulations examining dislocation interactions with He bubbles and voids have demonstrated that voids are stronger obstacles to dislocation motion as compared with pressurized He bubbles containing He-to-vacancy ratios less than 2, where loop punching is not prevalent [96-99]. Therefore, it follows that this transition in cavity pressure with fluence, which can also be considered a transition from bubble to void defect microstructures, could be related to the increase in hardness due to the formation of a higher density of stronger obstacles impeding dislocation motion.

Softening at the lower end of the fluence spectrum was present in both the AA-W and SPD-W microstructures implanted at 950°C. Given that these implantations were performed under identical conditions, the disparity in the degree of softening between the two samples implies that the microstructure is a contributing factor to this behavior. Prior MD simulations of He implanted microstructures have indicated that increased intragranular concentrations of He in the presence of grain boundary sinks can facilitate the coalescence of He bubbles within the boundaries, which in turn promote softening through a transition to grain boundary sliding and intergranular fracture [100]. Softening due to He bubble accumulation in grain boundaries has been observed in He implanted Cu [101] and W [22, 95]. Our results further demonstrate that He cavity induced softening transpires in W containing different microstructures, which influence the extent of this effect and logically follows from the notion that the presence of He bubbles at grain boundaries controls the degree of softening. A reduction in grain size is intrinsically accompanied by an increase in the grain boundary volume, which will act to further bias bubble formation to the grain boundaries, and in turn enhance the extent of softening under implantation conditions that concomitantly limit the density of intragranular bubbles. On this basis, the higher degree of softening in the AA-W appears counter-intuitive given AA-W's larger average grain size relative to SPD-W. However, as noted in Section 3, the AA-W microstructure is also highly textured with elongated grains parallel to the sample surface but nanocrystalline normal to the surface [65]. While the SPD-W also contains nanocrystalline grains, they are dispersed within an ultrafine grain size matrix. Thus, a nanoindentation measurement of hardness would probe a greater fraction of grain boundaries in AA-W, which is consistent with the reported higher degree of softening.

## 5. Strain softening due to bubble coalescence

Several behaviors common to the AA-W and SPD-W samples implanted at 950°C include softening at low fluence and subsequent hardening with an increase in fluence. While the latter is expected based on classical cavity induced hardening, the softening effect has been described in the context of prior observations from the literature and an assumed mechanism based largely on the differences in the degree of softening between the two W microstructures. We also note that cavity hardening was accompanied by a reduction in the modulus, which is consistent with an increased density of larger cavities within the grain matrix. Mechanistically, the influence of He defect distributions on plastic strain accumulation requires further bridging to the measured mechanical response. To explore the relationship between tensile properties and bubble distributions, three simulation cells were constructed though the combined MD-MC method described in Section 2.2 where the baseline W structure contained a normal distribution of nanocrystalline grains with an average size of 15 nm. The finer grain size of the simulation structures relative to the experimental materials was deliberate given the inherent limitations on simulation size. This same base structure was employed for both He concentrations to allow for a direct comparison with the pure W structure vis-à-vis grain size and grain boundary character distributions.

The cumulative bubble size distributions for the W-1%He and W-5%He structures are shown in Figure 9a with the simulation structures in the inset demonstrating that the He bubbles solely occupied grain boundary sites. While these concentrations were selected to emulate the effect of increasing fluence (corresponding to the 1 x $10^{16}$ ions/cm$^2$ and 1 x $10^{17}$ ions/cm$^2$ conditions as estimated by SRIM in Figure 1b), they also encompass the previously reported grain boundary embrittlement concentration based on the model given in Ref. [102] as $G_{He}^c = 6 \cdot \varepsilon_{surf}/(a \cdot n \cdot E_{He}^{sol})$, where $G_{He}^c$ is the critical concentration of He, $\varepsilon_{surf}$ is the surface energy of the material, $a$ is the average grain size, $n$ is the atomic density of the material, and $E_{He}^{sol}$ is the solution energy of the He atom at a substitutional site in the perfect lattice. Adopting values of these variables for W from the same reference using a 15 nm grain size, the critical He concentration for embrittlement is estimated to be 2.9 at.% and consistent with other literature findings [103, 104]. Additionally, the hydrostatic pressures of the bubbles as determined from the atomic stress tensor were approximately 6 GPa for W-1%He and 2.6 GPa for W-5%He while the corresponding Laplace pressures from Eq. 4 are 5.2 and 1.9 GPa, respectively. This indicates that the bubbles are marginally over-pressurized, which is common in MD simulations due to the non-equilibrium process employed for introducing He into the grain structures [36, 105, 106].

Nonetheless, MD still provides useful mechanistic insights into the deformation behavior despite the over-pressured state of the bubbles inhibiting direct replication of the experimental conditions.

Uniaxial tensile curves generated on the simulated nanocrystalline W microstructures are shown in Figure 9b with the corresponding 0.2% offset stress and Young's modulus tabulated in the inset for each structure. A drastic reduction in the modulus accompanied the increase in helium concentration from nominally 210 GPa in pure W and W-1%He to 165 GPa in W-5%He, corresponding to a ~23% reduction in the Young's modulus. Similar effects in the presence of He have been previously reported in simulations of W [107-109]. Experimental measurements on Cu [58], Ag [110], and W [22, 47, 95, 108, 111] have also been shown to align with theoretical calculations [112-114] and generally attributed to the increased open volume of He-induced cavities [111] with correlations between increased matrix porosity and He content [96]. The reduction in the elastic modulus for the W-5%He structure was accompanied by changes in the yielding behavior as reflected by the 20% drop in the 0.2% offset stress (5.39 GPa) relative to pure W and W-1%He (6.85 GPa, 6.59 GPa respectively). The flow response was also markedly impacted by increasing concentrations of He where the stress serrations in the pure W and W-1%He structures deriving from dislocation slip were far less prominent in the flow curve for the W-5%He structure. Collectively, the reduction in the effective yield point and transition in the flow behavior indicate that the increased fraction of grain boundary He bubbles in W-5%He promoted as shift to grain boundary plasticity.

Despite the vastly different strain rates and stress states, the reductions in strength and modulus from our simulations demonstrate a similar softening effect to the experiments. The change in flow behavior supporting enhanced grain boundary plasticity in the presence of grain boundary bubbles is also consistent with the experimentally determined changes in cavity size, circularity, and density suggestive of deformation-induced cavity coalescence. The mechanism was explored in our simulations by taking advantage of the ability to directly probe the evolution of grain boundary He bubbles. Bubble size distributions are mapped in Figure 10a at two different strains relative to the undeformed structures. The distributions in W-1%He aligned with the unstrained microstructure for strains up to 5% whereas a small subset of the bubbles in the tail of the distribution evolved to larger diameters in W-5%He. An increase in the applied strain to 10% produced conspicuous coarsening, as evidenced in the bubble volume in Figure 10b especially for the W-5%He structure, with a simultaneous reduction in bubble density in Figure 10c.

The combined changes in the bubble volume and density in W-5%He are indicative of bubble growth resulting from coalescence and consistent with the experimental trends. To understand the mechanism driving coalescence, a series of bubbles in the grain boundary plane at a strain of 4% are depicted in Figure 11a with atoms indexed based on their local value of dilatational stress. A region of high local stress is evident between the two bubbles, which upon an increase in strain to 5% in Figure 11b, coalesced with a corresponding relaxation of the dilatational stresses. This same region is shown from a different perspective at a slightly higher strain in Figure 11(c,d) where indexing of the bubbles via a surface mesh more clearly reveals the coalescence process. These observations indicate that local dilatational stress concentrations drive the coalescence of neighboring bubbles, which in turn relaxes bubble-induced stresses within the grain boundary plane.

Preferred grain boundary site occupation of He relates to its solution energy, which has been shown to be correlated with the site volume where He prefers to occupy sites of larger volume [115]. Local dilation due to the presence of He gaseous impurities promote the formation of stress concentrations and subsequent grain boundary structural transformations [116]. Such transitions redistribute the excess free volume in the grain boundary plane and therefore biases the diffusion of He to regions of elevated stress [117], such as shown for the regions between pre-existing He bubbles in Figure 11. It follows that the high stress concentrations in the vicinity of pre-existing He bubbles promote diffusion of He to regions with the highest dilatational stresses, providing a pathway to accommodating strain within the grain boundaries. This mechanism is apparent when comparing the mean squared displacement (MSD) of the grain boundary atoms following the effective yield point. At 3% strain, the MSD of the grain boundary atoms in the pure W, W-1%He, and W-5%He are 0.92 Å² (standard deviation, $\sigma$ = 1.26), 1.17 Å² ($\sigma$ = 1.46), and 1.22 Å² ($\sigma$ = 1.77), respectively. The grain boundaries in the W-5%He have therefore experienced more extensive local deformation due to the presence of a higher fraction of He bubbles, which in turn drives bubble coalescence to relax grain boundary stresses and suggests that the presence of He promotes a shift to grain boundary mediated deformation mechanisms.

## 6. Deformation mechanism shifts in the presence of coalesing grain boundary bubbles

The non-negligible MSD for W-5%He relative to W-1%He suggested a stronger contribution from the grain boundaries to strain accommodation even prior to the onset of plasticity. This is further supported by the suppressed flow serrations in the stress-strain curve for W-5%He relative to the other structures. Such stress serrations typically derive from strain localization events, one of which corresponds to the emission of dislocations from stress concentrations at interfaces in nanocrystalline materials [118-120]. Bubble induced grain boundary stress concentrations are profuse in the W-5%He structure. One such configuration at a strain of 4.1%, corresponding to the first discontinuity on the flow curve in Figure 9b, is shown in Figure 12a with the atoms colored based on their value of dilatational stress. A stress concentration is apparent at the intersection of two bubbles, from which a dislocation can be seen to nucleate in the microrotation indexed snapshot in Figure 12b. An increase in strain was accompanied by the dislocation traversing the entire grain before being absorbed at the adjacent grain boundary, and the resulting dislocation configuration following full slip is shown in Figure 12c at 4.2% strain. The atoms are deliberately indexed by their value of microrotation to reveal the dislocation structure, which for a full slip event, wouldn't be visible in either a common neighbor analysis (CNA) or centrosymmetry parameter. The dislocation emitted along the (110) slip plane with Burgers vector of a/2<111>, but then cross-slipped along the $(10\bar{1})$ slip plane to an adjacent (110) slip plane before being absorbed into the adjacent grain boundary.

Dislocation emission directly from He bubbles is widely discussed in the context of loop punching specifically in the BCC W lattice [36, 121-125] where W self-interstitial atoms (SIAs) are forced out by the internal pressure of the bubble via trap mutation [121] and orient to form <100> and <111> clusters. Rearrangement of these clusters leads to the formation of a stable prismatic dislocation loop with a Burgers vector of 1/2<111> [125], and recent findings have revealed that intermediate shear dislocation loops also contribute to the evolution process for larger bubbles [122]. Loop punching was also shown to be responsible for dislocation nucleation from a symmetric Σ3 grain boundary in Fe but accompanied by the formation of a disconnection that emitted a dislocation segment through the grain boundary plane [126]. From Figure 12, the highlighted dislocation nucleation event in the W-5% He structure did not emanate from the bubble surface; rather, it was emitted from a stress concentration ahead of the two denoted grain boundary bubbles, indicating that loop punching was not the governing mechanism for dislocation emission. As demonstrated in a small-angle Σ73b grain boundary in Fe, SIAs produced upon growth of a He cluster extended along the grain boundary plane due to its larger excess volume, which led to the He cluster adopting a longitudinal shape that aligned with the orientation of the extended grain boundary defect [126]. Atoms with large values of microrotation emanated from the intersection of the bubbles to the point of nucleation for the dislocation in Figure 12c. Combined with the presence of elongated grain boundary bubbles, these observations indicate that SIAs produced as a byproduct of He bubble growth clustered in a similar fashion favoring the excess free volume of the grain boundaries. Dilatational stress gradients consequently formed ahead of the bubbles that drove bubble coalescence (Figure 11) while serving to nucleate lattice dislocations at the intersection of the grain boundary plane with the BCC W matrix.

Strain softening in fine grained materials has indeed been correlated to the formation of stress concentrations at interfaces promoting strain relaxation through dislocation emission [120, 127]. The second localization event at 5% strain, which produced a more pronounced stress drop relative to the dislocation slip event at 4% strain, further substantiates the stress concentration mechanism based on the type of defect formed and its evolution with increasing strain. This sequence is shown in Figure 13(a-c) with atoms indexed at common strains via local coordination, microrotation, and dilatation stress, respectively. Just prior to the localization event at an applied strain of 5.2%, a stress concentration formed at the triple junction between three bubbles as indicated by the black arrow. An increase in strain to 5.3% was accompanied by nucleation of a planar defect from this stress concentration that remained in the lattice despite traversing the grain and impinging on the adjacent grain boundary. From the snapshots at larger strain, this defect thickened and contained a region of BCC coordinated atoms exhibiting large values of microrotation with boundaries between the surrounding BCC matrix. Collectively, these observations are consistent with the nucleation of a deformation twin [128], which is confirmed by the crystallographic indexing in Figure 13a and presence of coherent twin boundaries (CTBs) with a habit plane of {111}, or in coincidence site lattice indexing, Σ3 boundaries with a boundary plane of {112}. While the formation of twins in BCC metals is rare due to their high stacking fault energy [129, 130], unstable twin boundaries have been shown to form under extreme conditions such as high stresses, high strain rates (intrinsic to our MD deformation simulations), and low temperatures [131-133]. In our simulations, the observed twin nucleated at a large He bubble located a triple junction, which collectively act to enhance the local stresses as shown in the atomic dilatational stress maps. Stress concentrations under the high strain rate therefore control defect nucleation in structures containing He bubbles at grain boundaries while providing favorable conditions under certain instances for the nucleation of deformation twins.

Subsequent deformation produced continued strain localization and concomitant growth of the twin as evidenced in the CNA indexed snapshots in Figure 13a. Migration of the CTB transpired through a discrete process accommodated by localized shear displacements, which are captured in the microrotation indexing in Figure 13b and consistent with the CTBs providing an additional highly mobile glide plane through the grain. However, the individual steps across the CTBs in Figure 13b indicate that the CTBs are migrating via the motion of twinning disconnections, consistent with prior observations of twin boundary migration [134]. Disconnections were also observed in CTBs in MD simulations of BCC Fe – the Burgers vector for disconnections on the $(1\bar{1}2)$ deformation twin was found to be $\frac{1}{6}(111)$ with a magnitude of $\frac{\sqrt{3}}{6}a_0$ with a critical resolved shear stress of approximately 20 MPa, much smaller than the 82 MPa for the ½ <111>{110} edge dislocation [135]. The step height of this disconnection would be equivalent to the interplanar spacing between (112) planes, similar to those in Figure 13a. In nanocrystalline W, it was observed experimentally that deformation twins are often inclined with respect to the twinning plane, implying significant disconnection content. However, the high mobility of these steps on the deformation twins causes them to migrate rapidly, facilitating rapid detwinning [132]. The combination of this rapid detwinning process and their barrier to formation in W contributes to their near absence outside of in-situ microscopy, shock-loading, and high-rate MD simulations. Our results now add that the presence of He bubbles at grain boundaries combined with a high deformation rate can bias the formation of deformation twins, and even if short-lived, facilitate additional microplasticity in nanocrystalline W under extreme conditions.

Thus far, deformation behavior has considered instabilities deriving from localized slip events and their coupling to stress concentrations, which collectively acted to drive bubble coalescence and in turn enhanced strain localization through the activation of unstable mechanisms such as deformation twinning. Strain accommodation through the formation of twins in the presence of He grain boundary bubbles was accompanied by a reduction in grain boundary mediated dislocation activity as observed in the deformation snapshots in Figure 14(a-c). With planar defects and grain boundaries revealed by removing the BCC coordinated atoms and indexing the remaining population based on their value of shear strain, $\gamma$, the large fraction of dislocations present at both strains in the pure W sample in Figure 14a was noticeably reduced with increasing He concentration in Figure 14(b,c). The microrotation snapshots from 5.2-5.4% strain in Figure 13b indicate that local atomic shuffling and rearrangement within the grain boundaries were correlated with the bubble coalescence and influenced the nucleation of the deformation twin. Locally elevated values of microrotation also accompanied the dislocation emission event highlighted in Figure 12.

To quantify the effect of He bubbles on the degree of strain partitioning to the grain boundaries, the atomic shear strain was mapped for all grain boundary atoms as a function of applied strain using the contour plots in Figure 14(d,e) for pure W and W-5%He, respectively. A relative intensity scale was employed that normalized atom counts by the maximum frequency at the mode, which allows for direct comparison of the distributions across strains rather than emphasizing relative differences in the probability densities at each strain. Below an applied strain of ~2%, the shear strain distributions for the two structures were similar and unimodal but subsequently diverged as the strain was increased beyond the yield point and into the strain softening regime. The pure W structure retained a unimodal distribution across all strains with a large fraction of atoms exhibiting intermediate shear strains while the distributions bifurcated above 2% strain for W-5%He with elevated populations of grain boundary atoms exhibiting both larger and smaller shear strains relative to the values for pure W. With grain boundary atoms divided into different shear strain populations in Figure 14f, the He concentration dependence of this transition is captured by the simultaneous increase in the atomic fractions of the small ($\gamma \leq 0.1$) and large ($\gamma \geq 0.7$) shear strain populations and accompanying decrease in the intermediate ($0.1 < \gamma < 0.7$) population.

The bifurcation of shear strain is a manifestation of strain localization in the grain boundaries where a small population of atoms accommodate plastic strain, leaving the surrounding atoms nominally undeformed in the extreme case of plasticity in disordered solids [136]. In nanocrystalline systems, strain localizes in grain boundaries often facilitating other deformation processes such as grain boundary sliding, grain rotation, and the nucleation of lattice defects [137-140]. Our results demonstrate that the presence of He bubbles also promotes localized grain boundary plasticity in nanocrystalline W but with the partitioning of strain during localization less severe relative to an amorphous solid due to the coupling with complementary deformation mechanisms. Therefore, stress assisted He bubble growth and coalescence at grain boundaries in W leads to a shift towards interface mediated plasticity and potential activation of unstable deformation mechanisms (i.e., deformation twins), both of which have been correlated to softening in nanocrystalline systems. In the samples implanted at the lowest fluence at 950°C, cavity formation was biased to grain boundaries and coalesced during deformation with concomitant softening. These observations are consistent with the shift to enhanced interfacial plasticity deriving from stress assisted bubble coalescence in our simulations. The absence of softening at higher fluences was due to the increased matrix cavity size and density promoting a shift in the dominant deformation mechanisms from interface mediated to classical lattice cavity and loop induced hardening.

7. **Conclusions**

The influence of He cavity microstructures on the scaling of hardness and complex modulus was quantified as a function of fluence through targeted implantations at two temperatures in two grades of ultrafine-grained tungsten – AA-W containing elongated ultrafine grains with a {001}<110> fiber texture and SPD-W with a bimodal distribution of nanocrystalline and ultrafine grains of random texture. Nanoindentation property mapping and mapping of bubble morphologies within the plastic zone revealed:

- At 950°C, above the threshold for preferential grain boundary cavity formation, both microstructures exhibited reductions in hardness relative to the pristine condition at the lower end of the fluence spectrum where intragranular cavities remained sufficiently small to limit classical hardening effects.
- The higher degree of softening in AA-W was consistent with the nanoindentation measurements probing a larger fraction of grain boundaries due to the nanocrystalline length scales of the elongated surface grains in the direction of the expanding plastic zone.

- Collective changes in the mean cavity size, density, and circularity in the deformed region of the microstructure on an implanted surface indicated that cavity coalescence accompanied mechanical softening.

To build a mechanistic understanding of this behavior, MD simulations of polycrystalline tungsten structures containing grain boundary He bubbles (average concentrations corresponding roughly to two fluence conditions from experiments) were performed and shown to exhibit reductions in the flow stress akin to the experimental softening effect. By tracking the density and size of the grain boundary bubbles as a function of applied strain with correlative mapping of the local dilatational stresses, bubble coarsening was shown to transpire through a stress-assisted coalescence process. The resulting shift to larger bubble size distributions promotes strain localization in the grain boundaries through local atomic shuffling and sliding, grain boundary dislocation emission, and the nucleation of unstable defects such as deformation twins. Softening is attributed to this grain boundary mediated plasticity transition in the presence of He bubbles with the activation of unstable deformation mechanisms, even if short-lived, facilitating additional microplasticity in nanocrystalline tungsten under extreme conditions. While tailored nanostructuring continues to be a pathway for developing sink engineered alloys with promise for advancing tungsten as a first-wall material, the biased distribution of He at grain boundaries and its implications for the transition to unstable grain boundary-mediated plasticity can have a significant impact on tungsten's mechanical behavior in the aggressive fusion environment.

**Acknowledgments**

This work was supported by the National Science Foundation under Award 1810040. The authors would like to acknowledge Dr. Jonathan Gentile for his assistance with the nanoindentation experiments and Matthew Chancey for his support on the 500 keV He

**Figures**

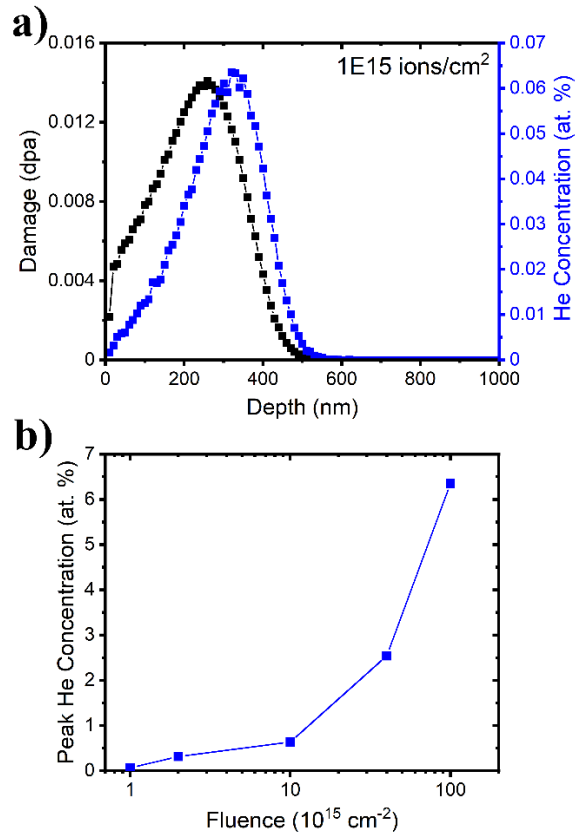

**Figure 1.** (a) Damage dose and He concentration as a function of depth in W generated from SRIM calculations [69] for 150 keV He$^+$ ions implanted to a fluence of $1 \times 10^{15}$ ions/cm$^2$. (b) Peak He concentration as a function of fluence over the range considered in our experiments.

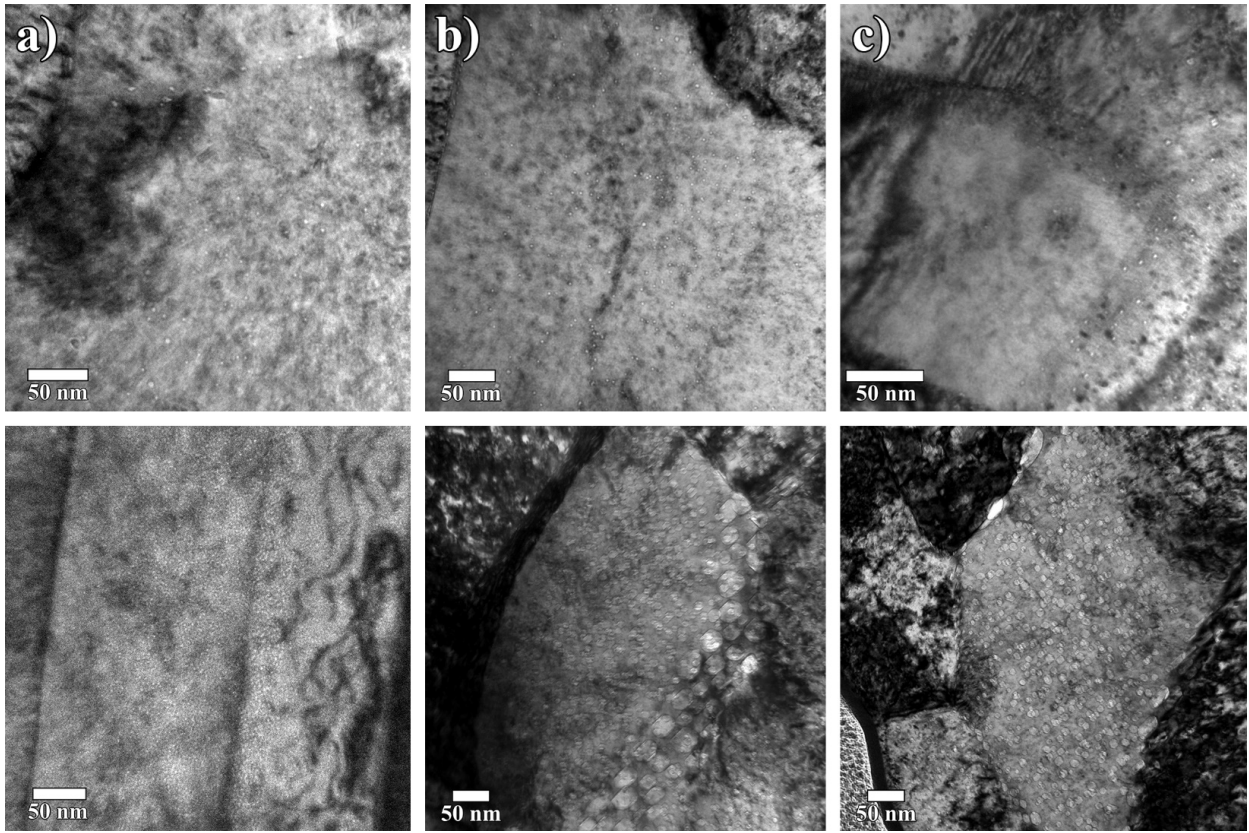

**Figure 2.** Bright-field TEM micrographs of AA-W implanted with (a) 500 keV He$^+$ ions at 800°C to fluences of (top) 1 x 10$^{15}$ ions/cm$^2$ and (bottom) 4 x 10$^{16}$ ions/cm$^2$ and (b) 150 keV He$^+$ ions at 950°C to fluences of (top) 1 x 10$^{15}$ ions/cm$^2$ and (bottom) 1 x 10$^{17}$ ions/cm$^2$. (c) Bright-field TEM micrographs of SPD-W implanted with 150 keV He$^+$ ions at 950°C to fluences of (top) 1 x 10$^{15}$ ions/cm$^2$ and (bottom) 1 x 10$^{17}$ ions/cm$^2$.

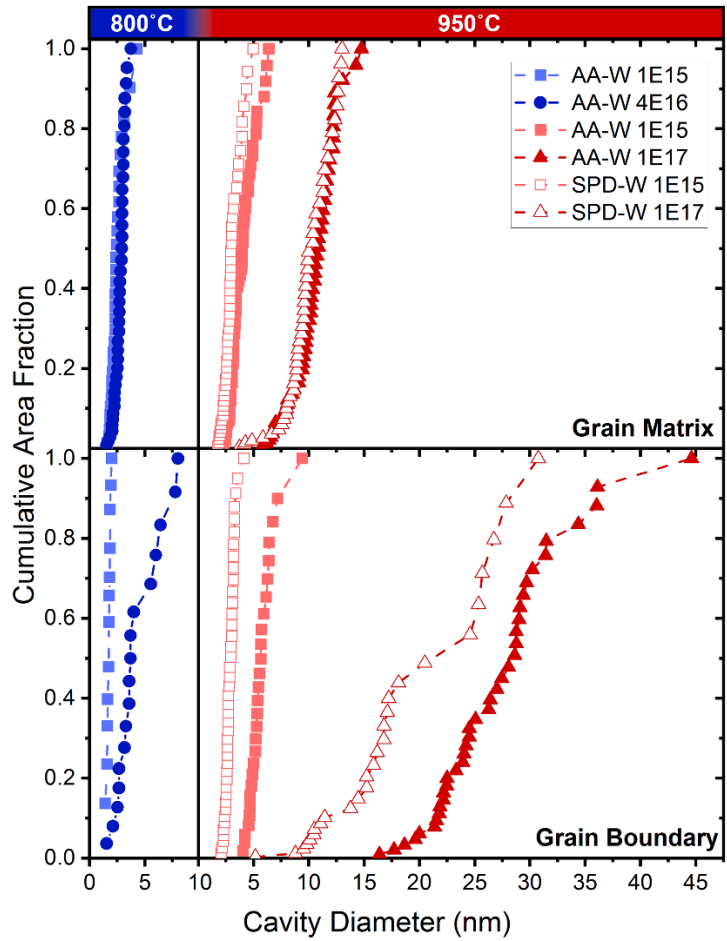

**Figure 3.** Cumulative cavity size distributions in the grain matrix (top) and at the grain boundary (bottom) for both microstructures grouped based on temperature for the different fluence conditions.

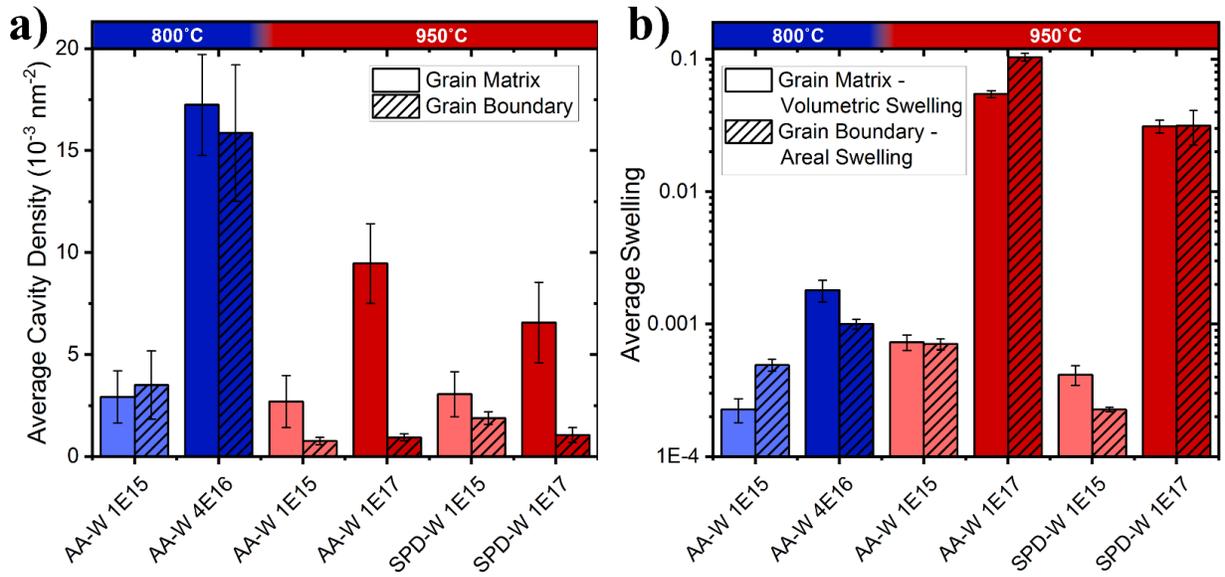

**Figure 4.** (a) Average cavity density and (b) average volumetric swelling delineated for the grain matrix and boundary at each fluence in the respective sample (AA-W and SPD-W). Data is presented in increasing fluence order for each tungsten microstructure and colored based on the implantation temperature.

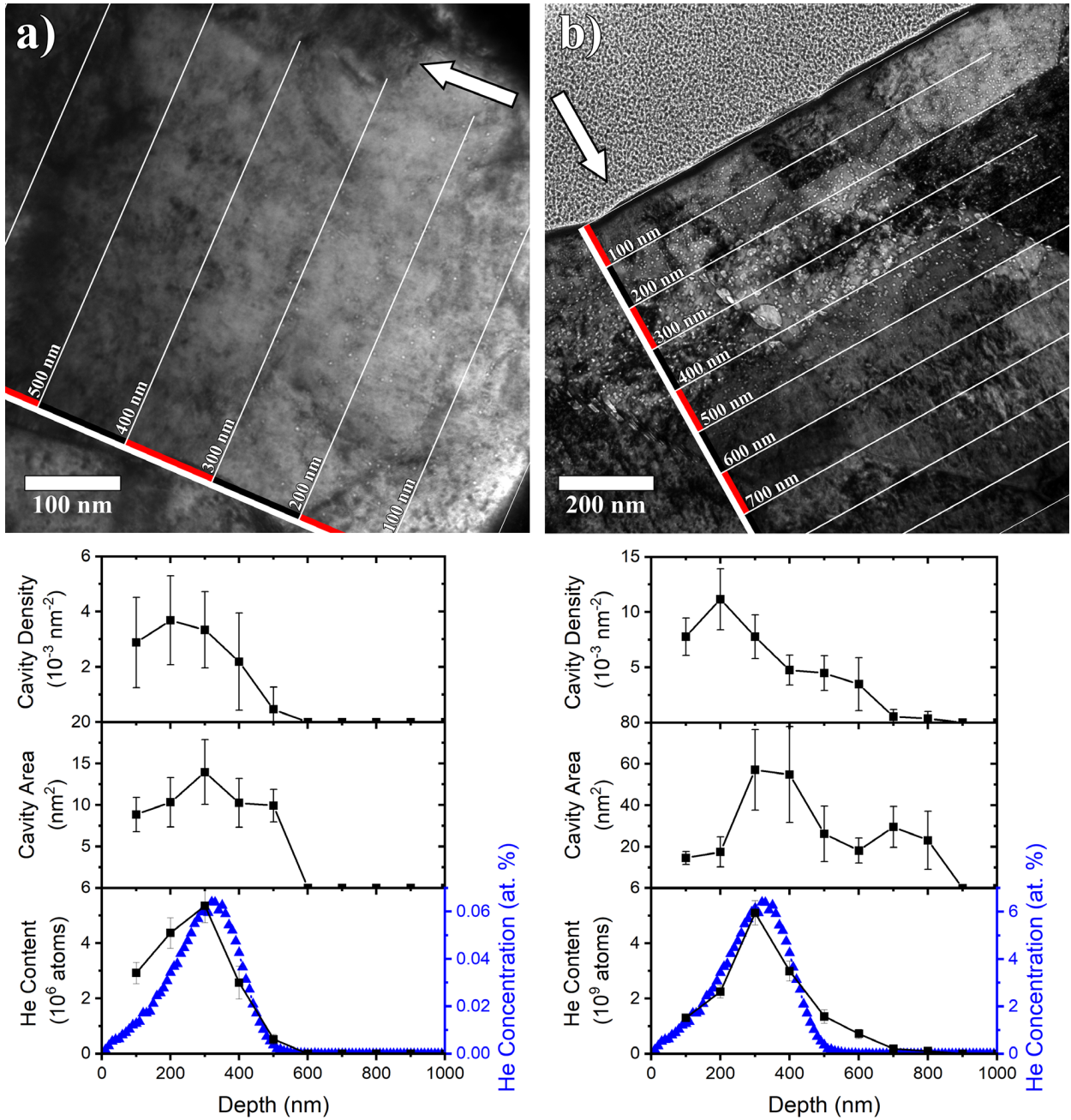

**Figure 5.** Average cavity densities, sizes, and estimated He content as a function of depth indexed as depicted on each microstructure for SPD-W irradiated with 150 keV He$^+$ ions at 950°C to fluences of (a) 1 x 10$^{15}$ and (b) 1 x 10$^{17}$ ions/cm$^2$. Grid lines map the depth from the sample surface with the white arrows indicating the direction of implantation. The blue concentration trends represent the He profiles from SRIM.

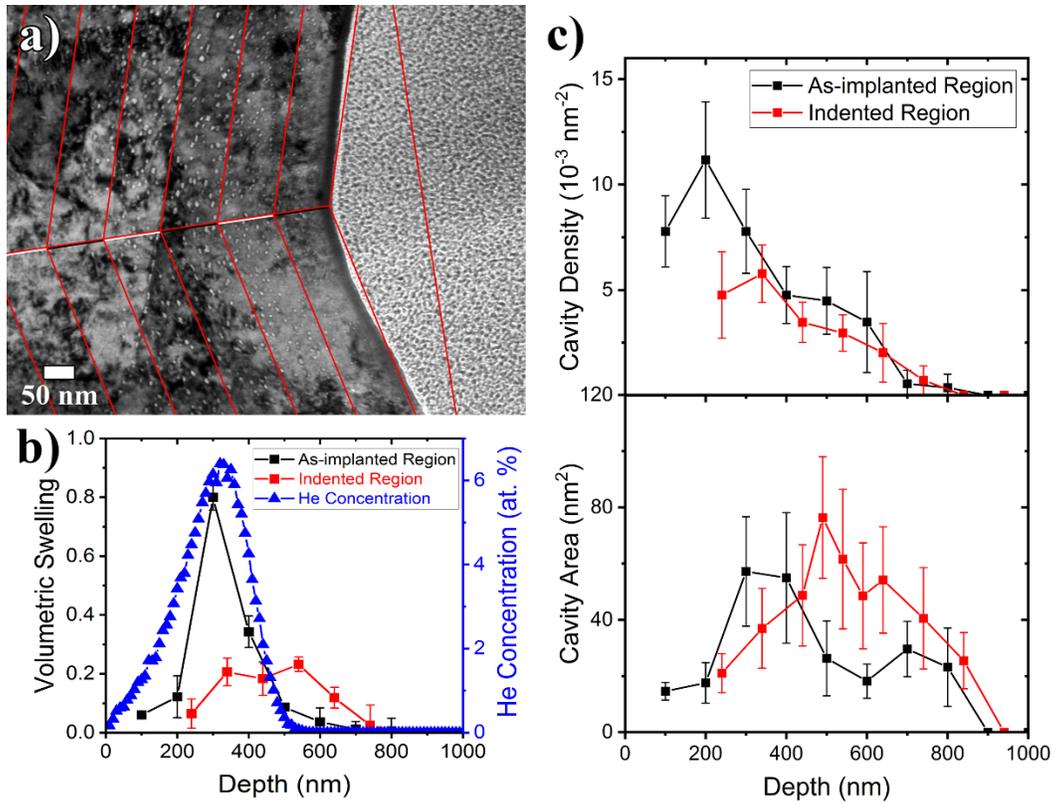

**Figure 6.** (a) Bright-field TEM micrograph of a cross-section through a residual impression on SPD-W irradiated with 150 keV He$^+$ ions at 950°C to a fluence of 1 x 10$^{17}$ ions/cm$^2$. (b) Volumetric swelling as a function of depth beneath the indent as compared with the profile for the as-implanted surface; the He concentration profile from SRIM is included for reference. (c) Average cavity density (top) and size (bottom) as a function of depth from the indent relative to the profile beneath the as-implanted surface.

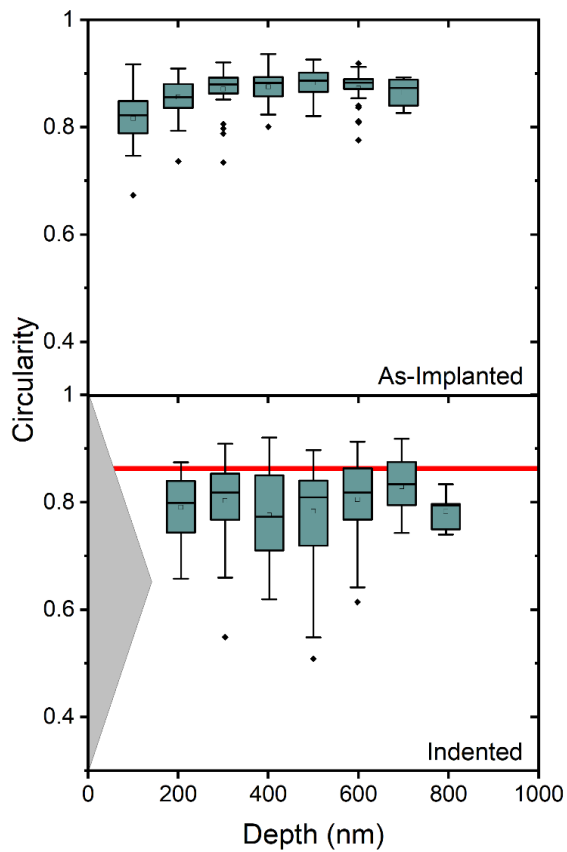

**Figure 7.** Cavity circularity in the as-implanted (top) and deformed regions (bottom) of SPD-W irradiated with 150 keV He$^+$ ions at 950°C to a fluence of 1 x 10$^{17}$ ions/cm$^2$. The inscribed red line denotes the average circularity for the as-implanted cavities as determined from the data in the upper plot.

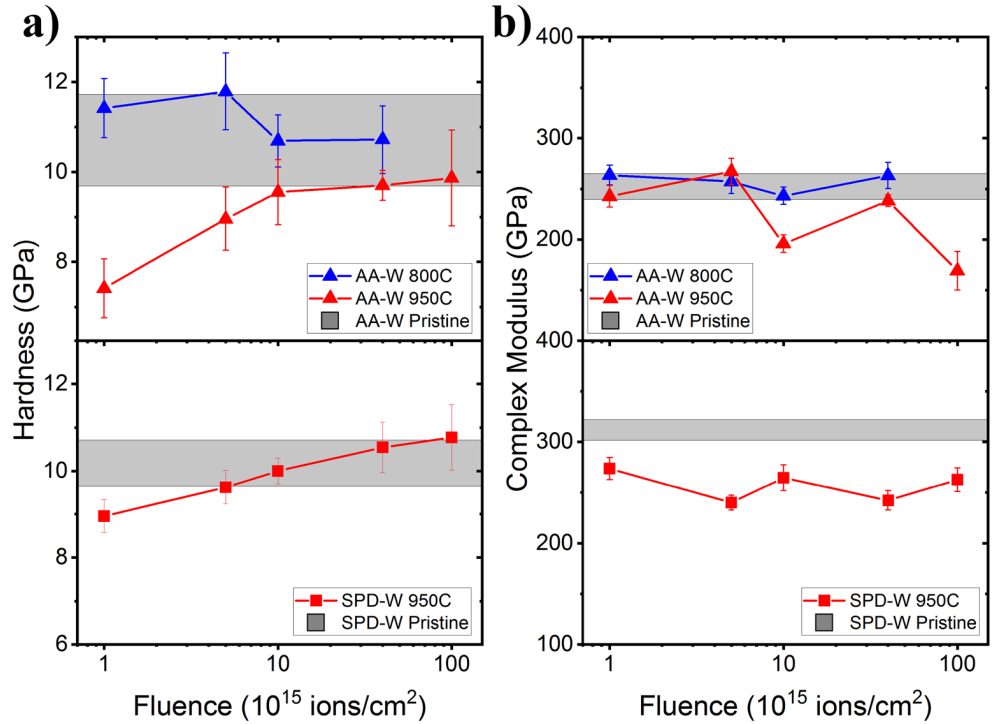

**Figure 8.** Hardness (a) and complex modulus (b) as a function of fluence for (top) AA-W irradiated with 150 keV He$^+$ ions at 950°C and 500 keV He$^+$ ions at 800°C, and (bottom) SPD-W irradiated with 150 keV He$^+$ ions at 950°C. Ranges for the hardness and complex modulus of the pristine annealed AA-W and SPD-W samples are represented by the grey regions.

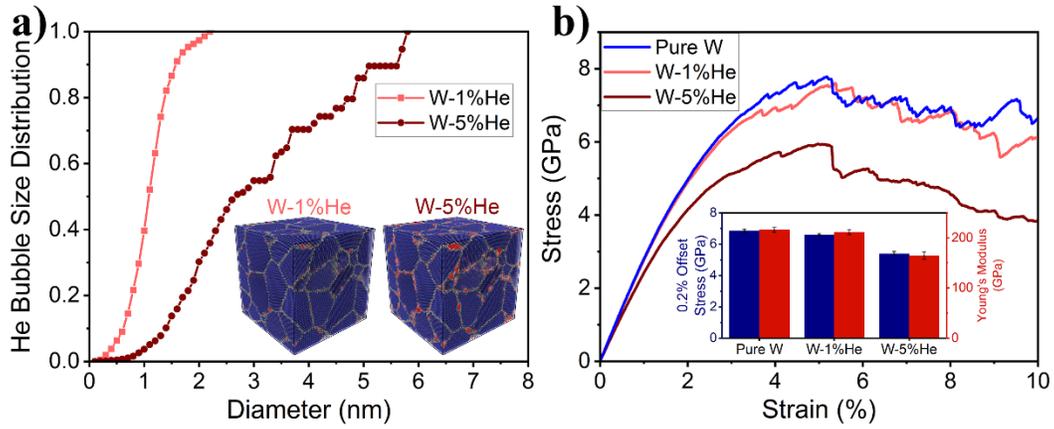

**Figure 9.** (a) Cumulative size distributions for the He bubbles formed in the simulated nanocrystalline W for He concentrations of 1% and 5% with the respective grain structures revealing preferential bubble formation in the grain boundaries (atoms colored based on their coordination with BCC atoms blue, grain boundary atoms grey, and He atoms red). (b) Simulated tensile curves for W, W-1%He, and W-5%He with the values for Young's modulus and the 0.2% offset stress provided in the inset.

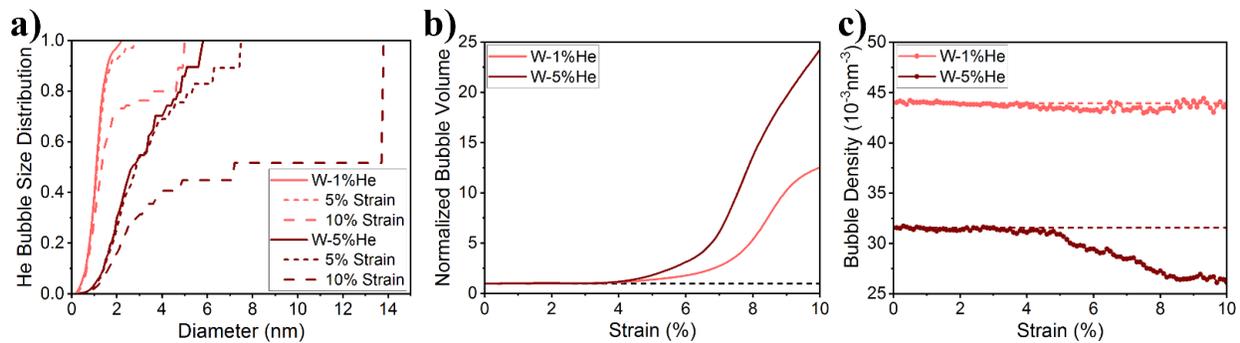

**Figure 10.** (a) He bubble size distributions in the simulated W-1%He and W-5%He structures in both the unstrained condition and at strains of 5 and 10%. (b) Normalized bubble volume and (c) density for the W-1%He and W-5%He structures as a function of strain.

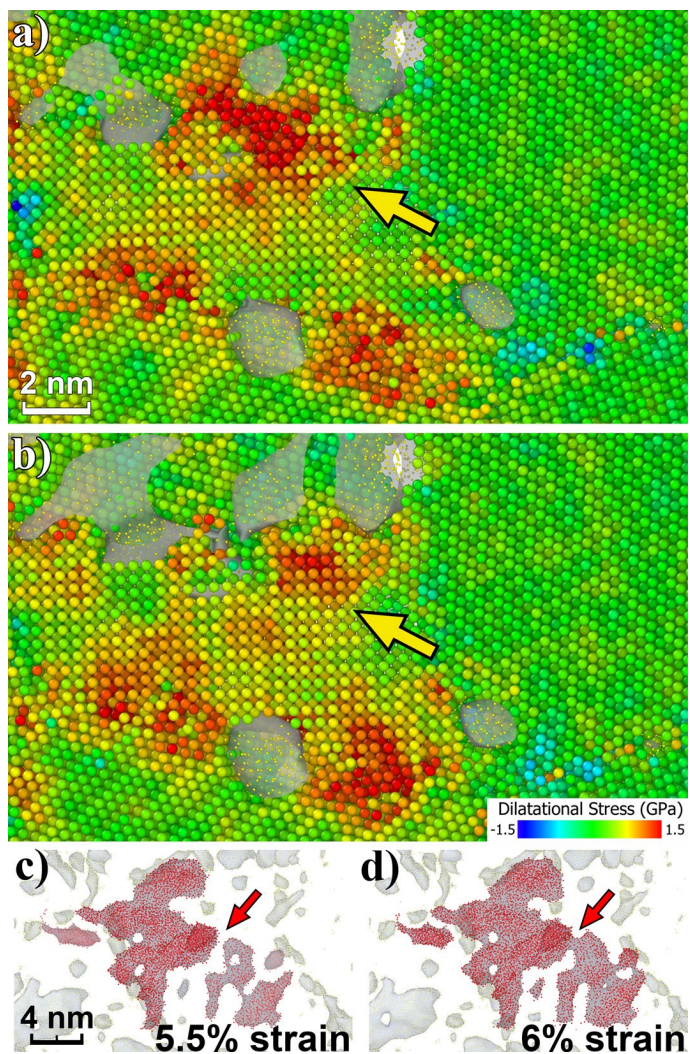

**Figure 11.** Deformation snapshots with atoms colored based on their dilatational stress at (a) 4.5% strain and (b) 5% strain with bubbles indexed via a surface mesh. The yellow arrows indicate regions of stress relief ahead of coalescing bubbles, which is shown at strains of (c) 5.5% and (d) 6.0%. The He bubbles that undergo coalescence are colored different shades of red representing before and after the coalescence event identified by the red arrows.

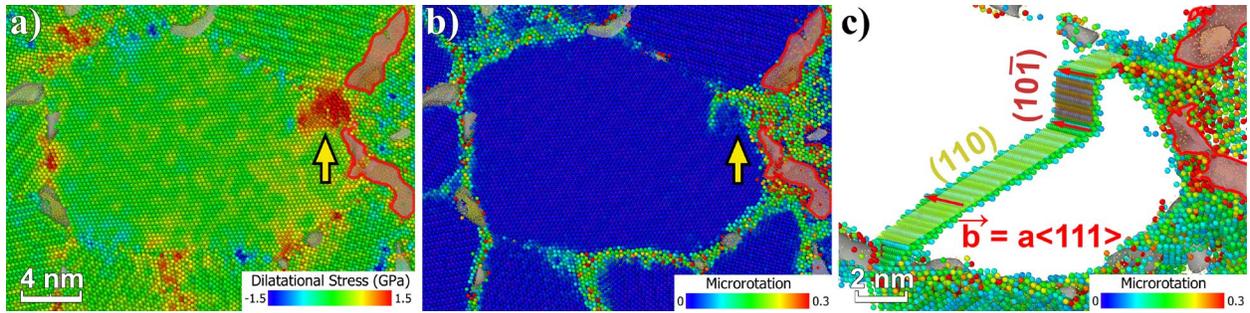

**Figure 12.** Deformation snapshots of the W-5%He structure at 4.1% strain with atoms colored according to their (a) dilatational stress and (b) microrotation. The yellow arrows highlight a dislocation nucleation event in (a) that emitted from the stress concentration in (b) at the intersection of the two bubbles (outlined in red) with the grain boundary as viewed parallel to the slip plane. (c) Snapshot of the full dislocation at 4.2% strain revealing a Burgers vector of a<111> with a (110) principal slip plane and cross-slip event along the $(10\bar{1})$ plane; atoms are colored based on microrotation.

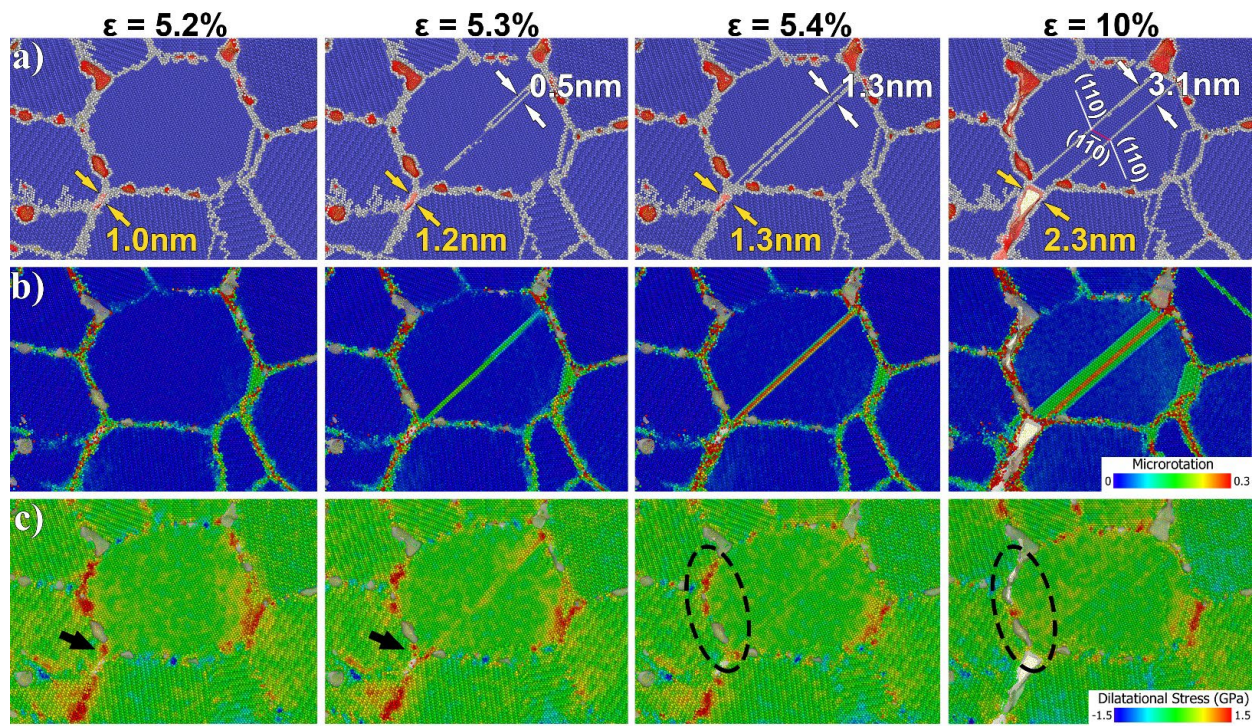

**Figure 13.** Deformation snapshots of the W-5%He structure with atoms colored according to (a) CNA, (b) microrotation, and (c) dilatational stress across four values of strain as indicated. The evolving width of the grain boundary bubble is inscribed by the yellow text in (a) with a large increase upon transition to a void at $\varepsilon = 10\%$. The width and crystallographic indexing of the deformation twin are shown by the white text in (a). The black arrows in (c) identify the stress concentration formed between the two bubbles where the deformation twin nucleated with the dashed circles highlighting stress release as the grain locally rotates upon subsequent growth of the twin.

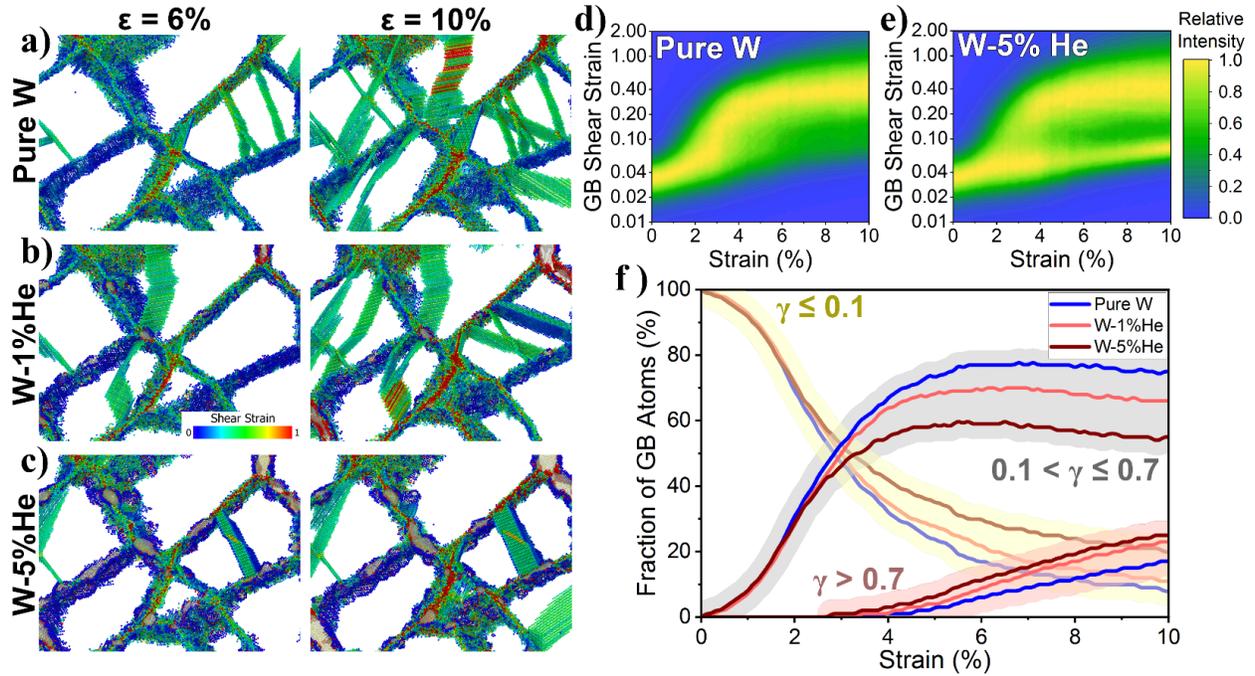

**Figure 14.** Deformation snapshots at 6 and 10% strain with intragranular BCC atoms removed and atoms occupying the grain boundaries and other defects colored by shear strain for (a) pure W, (b) W-1%He, and (c) W-5%He. Contour plots of capturing the grain boundary shear strain distributions as a function of applied strain for (d) pure W and (e) W-5%He; the distributions are normalized by maximum frequency at the mode. (f) Fraction of grain boundary atoms falling within the inscribed shear strain ranges as a function of applied strain.